# Imaging with high Dynamic using an Ionization Chamber[+]

Ralf -Hendrik Menk[a*], Heinz Amenitsch[b], Fulvia Arfelli[c], Sigrid Bernstorff[a], Hans Juergen Besch[d], Francesco Voltolina[c]

[a]Sincrotrone Trieste, S.S.14, km 163.5, Basovizza, 34012 Trieste, Italy
[b]Institute of Biophysics and X-Ray Structure Research, Graz, Austria
[c]Univ. degli Studi di Trieste, Trieste, Italy
[d]Fachbereich Physik, Universität-GH Siegen, 57068 Siegen, Germany

**ABSTRACT**

In this work a combination of an ionization chamber with one-dimensional spatial resolution and a MicroCAT structure will be presented. The combination between gas gain operations and integrating front-end electronics yields a dynamic range as high as eight to nine orders of magnitude. Therefore this device is well suitable for medical imaging or applications such as small angle x-ray scattering, where the requirements on the dynamic of the detector are exceptional high. Basically the described detector is an ionization chamber adapted to fan beam geometry with an active area of 192 cm and a pitch of the anode strips of 150 micrometer. In the vertical direction beams as high as 10 mm can be accepted. Every read-out strip is connected to an analogue integrating electronics channel realized in a custom made VLSI chip. A MicroCAT structure utilized as a shielding grid enables frame rates as high as 10kHz. The high dynamic range observed stems from the fact that the MicroCAT enables active electron amplification in the gas. Thus a single photon resolution can be obtained for low photon fluxes even with the integrating electronics. The specialty of this device is that for each photon flux the gas amplification can be adjusted in such a fashion that the maximum DQE value is achieved.

## 1. INTRODUCTION

The improvements achieved at modern synchrotron light sources and the associated optics in the last decade result in a tremendous increase in brilliance and subsequently photon flux, which was not, accompanied by an equivalent development of the x-ray detection devices. It is therefore very often the x-ray detector that limits the final data quality[1] no matter if single photon counters or integrating devices are used. Clearly, the use of single photon counters carry several advantages since they do not contribute additional noise and hence their dynamic range is almost unlimited at least as long no dead time loses are introduced. However, these devices are in general rate limited. On the other side integrating detectors have a higher rate capability but since they comprise always an inherent noise level they are not so suitable for low flux applications. In the following a novel detector device will be presented that comprises the advantaged of both detector types. Basically it is an integrating device based on the principle of a highly segmented gaseous ionization chamber, which can be operated in gas amplification mode. In this fashion the gas gain mechanism is used to adjust the total charge in the detector according to the incoming photon flux, so that for each single photon a signal is always generated which is significantly higher than the noise background of the integrating electronics. Hence, for almost all photon fluxes single photon precision can be obtained.

[+] Project is supported by the European Community Contract No. FMBICT 980104

[*] Correspondence: Email: ralf.menk@elettra.trieste.it; Telephone: +39 040 375 8201; Fax: +39 040 9080902
    or: Email: francesco.voltolina@ieee.org

## 2. DETECTOR SET-UP

### 2.1. Principle of charge generation and mechanic set-up

The one dimensional (see Fig. 1) detector [2] -is a highly segmented ionization chamber featuring an active area of 192 cm * 10 mm. In the configuration as described here it is suitable for the detection of x-rays in the energy range between (5-25 keV). As depicted in (see Fig. 2), photons entering the detector through the entrance window (thin carbon fiber) and the drift cathode (aluminized Mylar foil) are absorbed mainly by a photoelectric absorption in the detector gas (such as Ar / $CH_4$, Ar / $CO_2$ or Xe/$CO_2$ under pressure up to 3 bars). During this process $n = E_\gamma/W_{ion}$ primary charges are generated, when $E_\gamma$ is the energy of the photon and $W_{ion}$ is the mean energy required to create a ion / electron pair in the detector gas. For the distance of 20 mm between drift cathode and MicroCAT and a photon energy below 12 keV the quantum efficiency (QE) can be calculated as 100% for a Xe/$CO_2$ gas mixture at 3 bar pressure. In the constant drift field of 20 mm length the ions are transported to the drift cathode, whilst the electrons are transported to the gas amplification structure (MicroCAT)[3]. The MicroCAT is a 55μm thick nickel foil perforated with micro-holes (116 μm diameter) in a hexagonal arrangement. The center to center pitch of the holes is 164 μm. At a distance of about 200 μm below the MicroCAT, which is supplied with negative high voltage (typically 500-1000 V), the virtual grounded anode strips are placed. Due to this configuration the electric field strongly increases in the vicinity of the micro-holes reaching an almost constant plateau between the MicroCAT and the anode (typically 1-80 kV/cm). In this fashion almost all electrons produced in the primary charge are feed through the micro holes of the MicroCAT (transparency ~ 100%) and are transferred on the anode (in contrast to the 'optical transparency of the MicroCAT which lies in the order of 50%). Thus the MicroCAT structure is utilized as a shielding grid, which enables frame rates as high as 10kHz. If the field is high enough secondary electrons are produced, which subsequently build an avalanche in the vicinity of the anode (gas gain)[4]. Depending on the applied fields this avalanche comprises up to some 10 to $10^5$ electrons per primary electron. However, in order to maintain a constant gas gain along the direction of the position encoding, the distance between MicroCAT and anode has to be kept in the order of some10 μm. For the active area given here the use frame-like spacer outside the active region is sufficient as long as the anode features sufficient planarity.

The position encoding in the one direction (192-cm) stems from the subdivision of the anode into 1280-readout strips having a pitch of 150 μm between two adjacent strips in the active area. This anode (see Fig. 3) is realized using ultra thin multilayer processes on a Kapton foil[1], which was subsequently glued on a very precise aluminum board of 10 mm thickness. Typically the resistance between two adjacent strips is $10^{13}$ Ω.

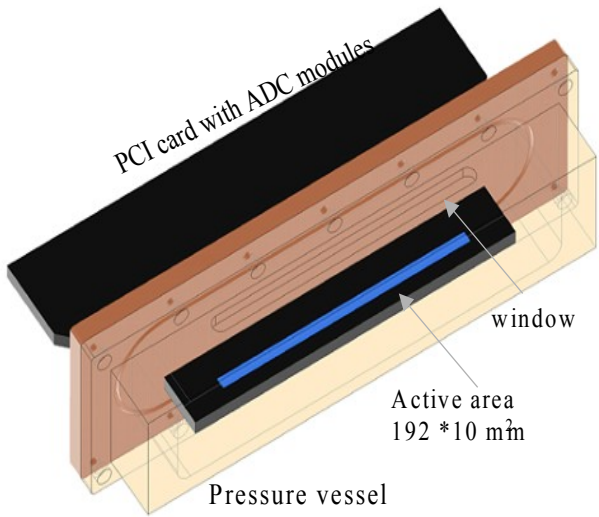 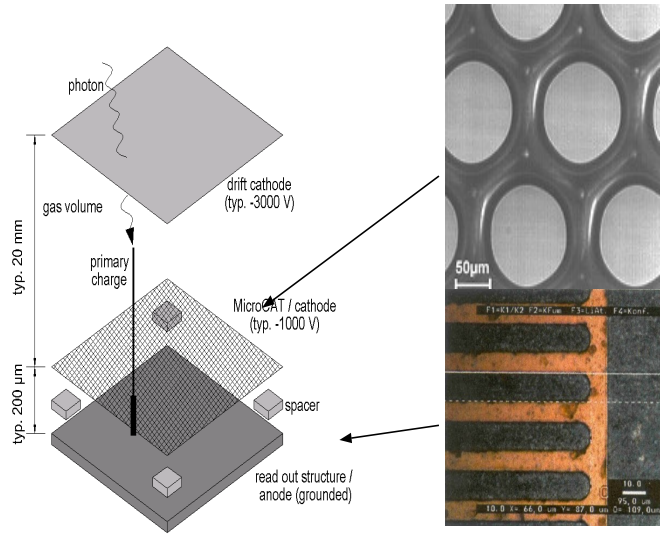

Fig. 1: Schematic view of the detector housing showing the read out anode bonded to the pressure vessel.

Fig. 2: Exposed view of the detector plus details of the gas gain structure (MicroCAT) and the position-encoding anode. The full active area of 192 x 10 mm² is composed of 1280 anode strips with a pitch of 150μm.

---

[1] ILFA Address, Lohweg 3, 30559 Hannover, Germany

As described before the precision in terms of planarity (better than 20 μm on 192 cm) of the board and thus of the read out anode is crucial for homogeneous gas gain operations. The anode is partially accommodated in an aluminum housing comprising the conversion gas mixtures mentioned before. It is sealed with the anode using Torr Seal$^{TM}$, which is an excellent sealant in vacuum equipment requiring vacuum down to $10^{-9}$ or lower. It forms a high strength bond with metals, ceramics or glass. Since no solvent are involved to evaporate this kind of glue ensures the purity of the gas as required for gas gain operations. The electrons released either in the primary charge generation (ionization chamber mode) or after the gas amplification (gas gain mode) are collected on the anode strips and integrated for a certain time (time frame) in attached amplifier channels.

## 2.2. Analog Electronics

Outside the pressure vessel the anode strips are fan-out and subsequently are connected via micro connectors to 10 VLSI carrier boards. Each carrier board comprises two custom made mixed analog digital chips that are successors of the JAMEX VLSI[5]. This new monolithic read out system (see Fig. 4) comprises 64 low noise analog integrators with reset switch followed by 64 sample and hold stages including a eight times correlated sampling stage as well as a 64:1 multiplexer and a differential output buffer. Sixteen different feedback capacities and subsequently 16 different gains can be programmed for eight blocks of eight channels. Those capacities range from 0.4pF to 47.4pF. Suitable integration times lay between 0.1 – 200 ms and can be selected by the user. However, the leakage current of the JFET gate, which builds the first amplifier stage, is of the order of some pA. For an integration time of 0.1 ms the equivalent noise charge accumulated is thus 25 electrons. This gives an acceptable signal to leakage current ratio considering that for instance a single 8 keV photon releases around 300 electrons (with no gain in the MicroCAT). Since the noise of the leakage current for that integration time is in the order of 5 electrons equivalent a unique discrimination for a single photon is ensured for an integration time of 0.1 ms and thus a frame rate of 10 kHz. While integrating the charges in the feed back capacity it is possible to read the contents of the sample and hold stages. However, due to the reset times of the feedback capacities a detector dead time or 8% at the maximum frame rate of 10 kHz is observed. This means a reduction in the detective quantum efficiency (DQE)[6] of 8%, which is for most applications acceptable.

## 2.3. Digital Electronics

The acquisition electronics is built around a commercially available multi channel ADC system developed by SUNDANCE[2] (see Fig. 5 and Fig. 6).
However, the Sundance system does not match the requirements of the Jamex Chips in the first place. Thus both systems have to be interfaced via a so-called Jamex to Sundance (J2S) interface card that is a big mixed-signal board 337x130 mm in dimension using 6 layers gold plated technology. As mentioned before the JAMEX features a differential current output stage to match the high requirements on the noise performances and rejection.

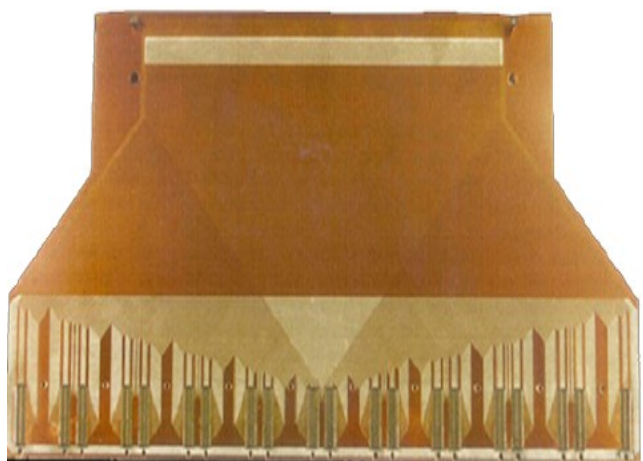

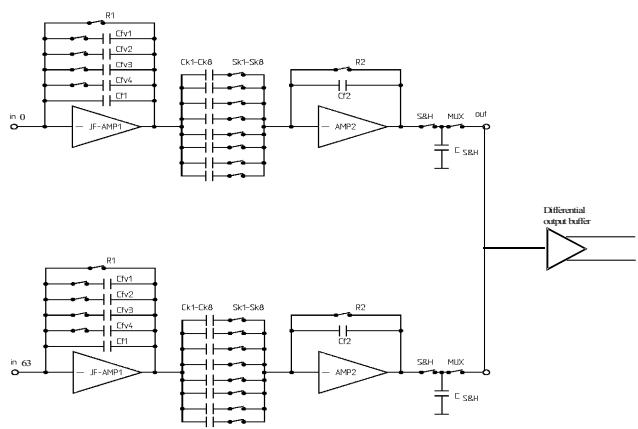

Fig. 3: Photograph of the readout anode.

Fig. 4: Sketch of 64 channel JAMEX monolithic read out system

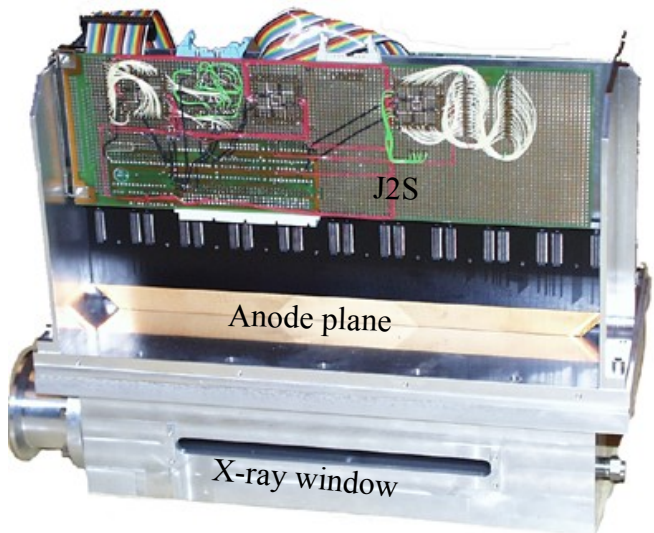 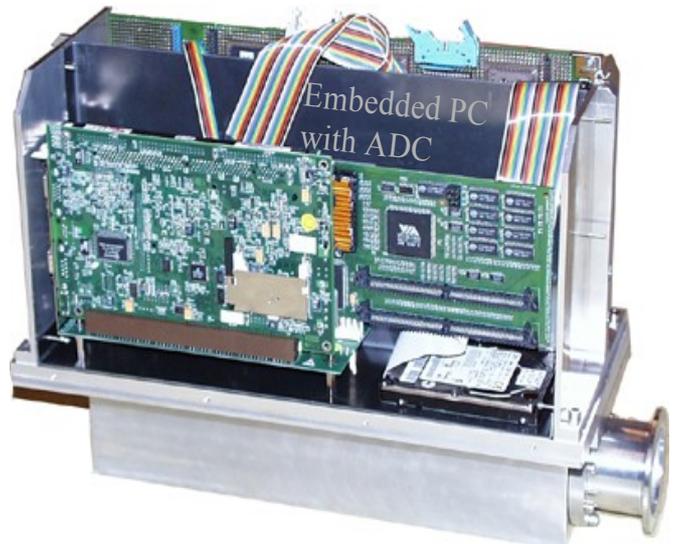

Fig. 5: Photograph of the front side of the detector. Visible are x-ray entrance window, partially the anode structure as well as the J2S.

Fig. 6: Backside of the detectors with embedded PC and carrier support for the multi channel ADC cards.

Therefore the J2S board performs differential to single ended conversion, individually programmable offset shifting in order to use full resolution of each A/D converter. Moreover an analog 1st order low pass filtering is provided in order to reduce the overall system noise of the chain. A PGA (programmable gain amplifier) provides better flexibility for high dynamic range imaging and a built in limiting function helps to avoid A/D overflow recovery problems. Another important function of the J2S board is to provide linear power supply, reference rails and the correct LVDS digital signals to control the operation and time frames of the 20 Jamex chips, allowing functionality like individual gain setting for each channel group, and different operation modes. The cabling between the Jamex hybrids (using flexifoil links) and the Sundance ADCs cards take place through several connectors on this separate board. Test signal generation capability is also embedded on the J2S board in order to give the possibility to test all the detector electronics without the need to use an X-Ray source.

On the SUNDANCE side a single PCI bus board carries (see Fig. 7) up to 4 standard TIM modules. In the present configuration 3 Sundance ADC modules (see Fig. 8) are used providing 8 ADCs @10 Mhz, 14 bit each, for a total of 24 analog inputs on 1 board. In this way it is possible to connect the outputs of the Jamex chips directly to 20 of the analog ADCs inputs available. The spare 4 inputs are used to monitor other signals and are used for debugging purposes. Nevertheless, since the output signals of the Jamex are balanced (differential) and the ADCs require single ended input an appropriate conversion on the J2S is needed. There are 2 additional modules on a second PCI carrier board containing one 320C64x DSP chip as well as one Virtex FPGA module. Next to other tasks the DSP is in charge for offset corrections and channel gain normalization. The DSP as well as the FPGA card are programmed using special communication ports on the PCI carrier board. Both PCI carrier boards are stacked inside the detector together with a small half-size PCI card embedded computer and a custom bus back-plane. All digital equipment is placed inside the detector in a specific shielding box in order to avoid electro-magnetic pick up by the Jamex chips or the buffering electronics on the J2S card. This configuration allows a very short cabling between the Jamex out put signal and the ADC inputs and thus improves in signal to noise and distortion in comparison with a long cable analog signal transmission solution.

In this solution the embedded PC communicated with the external world via standard Ethernet interfacing and allows therefore easy software maintenance and firmware reconfiguration for the whole system.

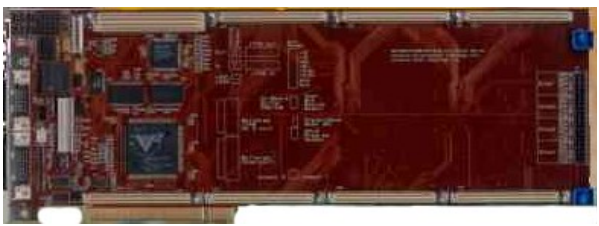 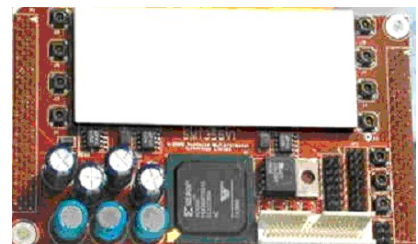

Fig. 7: Photograph of the Sundance PCI carrier board.

Fig. 8: Photograph of the Sundance ADC board.

## 3. DETECTOR PERFORMANCE

### 3.1. Spatial resolution

Since the described detector is in most cases operated as an integrating one (but mainly with single photon resolution) the following effects contribute to the spatial resolution: i.) range of photoelectrons ii.) diffusion of the primary charge cloud iii.) fluorescence / range of Auger electrons and iv) the segmentation. Even though the nature of the shielding structures could have a potential influence on the spatial resolution it was not observed during measurements. For 8 keV photons in $ArCO_2$ at 2bars and a drift length of 20 mm the range of photoelectrons and the diffusion contribute with 100 µm which is slightly better than the pitch of the segmentation. Only for photon fluences noticeable below 10 kHz per pixel and second a centroid method can be applied that substantially increases the position resolution. In this case it must be ensured that at maximum one photon registered at a time per pixel and integration interval.

For the measurement of the position resolution the detector was filled with an $Ar/CO_2$ gas mixture (90:10) at a pressure of 2 bar. Only two Jamex chips comprising 128 channels were attached to the detector system. Half of the active area was covered by a sharp edge while 8 keV x-rays illuminated homogeneously the active area. Several hundred time frames were averaged in order to decrease the statistical errors. The result is an intensity profile very similar to an edge function. A numerical derivation applied to this edge function delivers the so-called point spread function (see Fig. 9), which is the response function of the detector to a Dirac peak like stimulation. The FWHM of the Gaussian shaped PSF is 160µm, which is in agreement to simulations taking into consideration the four effects discussed above. The modulo of the Fourier transform is the modulation transfer function (MTF) that describes how good details of a certain size or equivalent a certain 'spatial frequency' are transmitted through the detection system. A common measure for the spatial resolution is the quotation of the 10 % level of the normalized MTF. For the system described here a resolution of 2.5 LP/mm can be quoted which is in good agreement with the theoretical limit of 0.16 mm (see Fig. 10).

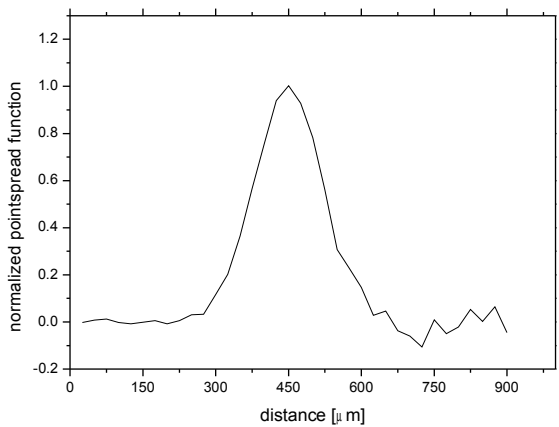 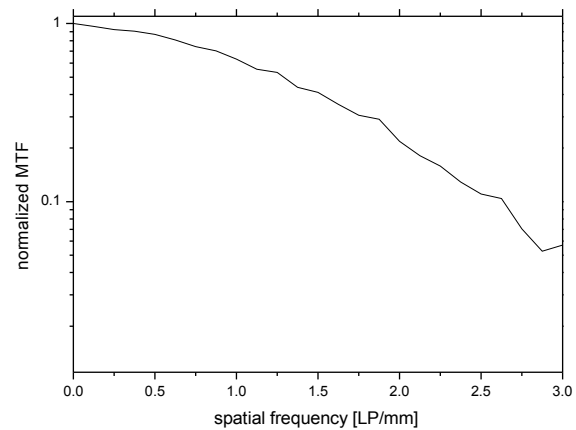

Fig. 9: Measured point spread function for 8 keV photons in a 2 bar $ArCO_2$ gas mixture.

Fig. 10: Calculated modulation transfer function for the conditions as quoted in Fig. 9.

### 3.2. Imaging / Diffraction measurements

Several preliminary experiments have been carried out at the SAXS beam line at the Sincrotrone Trieste (Italy)[7,8] in order to test the suitability of the detector in imaging and diffraction applications. As before also for these experiments the detector was equipped with two JAMEX chips only featuring 128 channels. Moreover, the detector was filled with an $ArCO_2$ gas mixture (90:10) at 2 bars. Beside direct beam measurements a dry rat-tail tendon collagen sample was used to generate a diffraction pattern to be imaged. In the community of small angle x-ray scattering experimentalists this sample is widely used as a standard for calibrating the instruments. A camera length of 77 cm was selected and the energy of the monochromatic primary beam was set to 8 keV. By placing different stacks of aluminum foils into the primary beam directly in front of the sample the photon flux could be controlled. At first the direct beam was imaged in ionization chamber mode (see Fig. 11). This two dimensional image was recorded while the detector was moved about 10 mm in horizontal direction through the x-ray beam. The maximum photon fluence was measured as $5*10^{11}$ photons /mm$^2$ /sec.

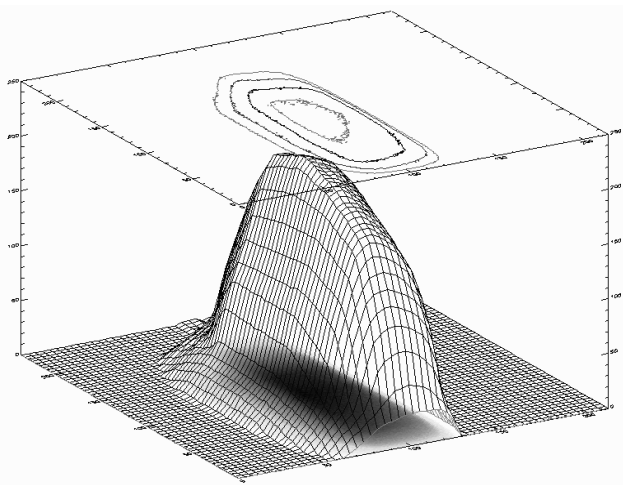 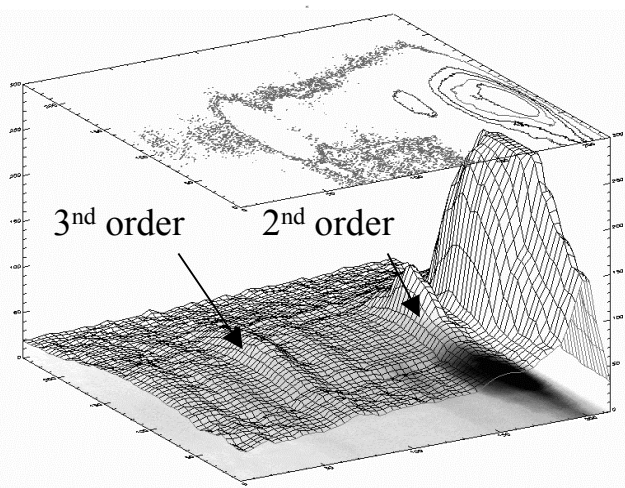

Fig. 11: Image of the direct beam of the SAXS beam line at the Sincrotrone Trieste recorded with the detector in ionization chamber mode featuring a fluence of 8 keV x-rays of $5*10^{11}$ photons /mm$^2$ /sec.

Fig. 12: Image of the diffraction pattern of a rat –tail tendon collagen recorded with the detector operated in gas gain mode with a gain of 1000. The 2$^{nd}$ and 3$^{rd}$ order reflex is clearly visible

For the next measurements the in-vacuum beam stop was placed behind the rat tail tendon collagen sample and the 2 dimensional diffraction pattern was recorded again by moving the detector in horizontal direction 10mm but this time applying gas gain. For a gas gain of 300 the output signal of the detector for that region was 34% of the entire ADC range. In addition the 2$^{nd}$ and the 3$^{rd}$ order reflex of the rat-tail tendon collagen diffraction pattern arose from the noise floor with a signal to noise ratio of 1.8. A further increase of the gas gain (up to 30000) increased the signal to noise ratio to 30 and above. This measurement is an excellent example how to increase the signal to noise for a given photon flux utilizing the gas gain in combination with the integrating electronics. Even for a single photon a signal can be generated which is significantly higher than the noise background of the integrating electronics.

## 4. CONCLUSION AND OUTLOOK

The unique feature of the Micro-CAT structure allows combining the advantages of a single photon counter with the characteristics of an integrating detector. If single photon resolution is required the gas gain can be adjusted in such a fashion that the charge released by the single photon is substantially higher than the equivalent noise charge of the integrating electronics. Moreover, the choice of the conversion gas and the gas pressure provides a high flexibility for applications in different energy ranges. A high intensity precision is combined with a good spatial resolution, a high local rate capability and time resolution of 100 μs.

The entire system with 1280 channels will be tested in May 2003 on the SAXS beam line.

## 5. ACKNOWLEDGMENTS

The authors would like to thank C. Fava from Sincrotrone Trieste for the design studies of the detector. The help of G.Schmid, R. Neumann of the mechanical workshop of the University Siegen is gratefully acknowledged. Moreover, the authors would also like to thank all of those who are not mentioned by name but whose contributions are appreciated. Eventually the research presented in this article has been supported by the European Community (contract no. FMBICT961694).